\documentclass[twocolumn,showpacs,preprintnumbers,amsmath,amssymb]{revtex4}

\usepackage{graphicx}
\usepackage{dcolumn}
\usepackage{bm}

\DeclareMathAlphabet{\mathpzc}{OT1}{pzc}{m}{it}
\DeclareMathOperator{\arctanh}{arctanh}

\begin{document}

\title{Anisotropic strange star in Finsler geometry}

\author{Sourav Roy Chowdhury}
\email{sourav.rs2016@physics.iiests.ac.in}
\affiliation{Department of Physics, Indian Institute of Engineering Science and Technology,
	Shibpur, Howrah 711103, West Bengal, India}

\author{Debabrata Deb}
\email{ddeb.rs2016@physics.iiests.ac.in}
\affiliation{Department of Physics, Indian Institute of Engineering Science and Technology,
	Shibpur, Howrah 711103, West Bengal, India}

\author{Farook Rahaman}
\email{rahaman@associates.iucaa.in}
\affiliation{Department of Mathematics, Jadavpur University, Kolkata 700032, West Bengal, India}

\author{Saibal Ray}
\email{saibal@associates.iucaa.in}
\affiliation{Department of Physics, Government College of Engineering and Ceramic Technology, Kolkata 700010, West Bengal, India  
\& Department of Natural Sciences, Maulana Abul Kalam Azad University of Technology, Haringhata 741249, West Bengal, India}

\author{B.K. Guha}
\email{bkg@physics.iiests.ac.in}
\affiliation{Department of Physics, Indian Institute of Engineering Science and Technology,
	Shibpur, Howrah 711103, West Bengal, India}

\date{\today}
             
\begin{abstract}
In the present paper, we report on a study of the anisotropic strange stars under Finsler geometry. Keeping in mind that Finsler spacetime is not merely a generalization of Riemannian geometry rather the main idea is the projectivized tangent bundle of the manifold $\mathpzc{M}$, we have developed the respective field equations. Thereafter, we consider the strange quark distribution inside the stellar system followed by the MIT bag model equation of state (EOS). To find out the stability and also the physical acceptability of the stellar configuration, we perform in detail some basic physical tests of the proposed model. The results of the testing show that the system is consistent with the Tolman-Oppenheimer-Volkoff (TOV) equation, Herrera cracking concept, different energy conditions and adiabatic index. One important result that we observe is that the anisotropic stress reaches to the maximum at the surface of the stellar configuration. We calculate (i) the maximum mass as well as corresponding radius, (ii) the central density of the strange stars for finite values of bag constant $B_g$ and (iii) the fractional binding energy of the system. This study shows that Finsler geometry is especially suitable to explain massive stellar systems.
\end{abstract}



\maketitle

\section{INTRODUCTION} 
In recent years, theoretical physics has been dealing with the properties of different spacetimes in four dimensions as well as in higher dimensions. It is interesting to note that most investigations are centred on spacetimes, typically solution to the Einstein field equations in supergravity theory. Obviously, there are various issues behind investigating gravitational theories which are more general than general relativity (GR). On the basis of quantum gravity researches, there are enormous theoretical predictions and these eventually indicate that GR should be superseded in more general aspect. As such apart from astrophysical significance, there are enormous applications of spacetime to string theory, ADS/CFT correspondence, de Sitter gauge theory, induced gravity~\cite{Banados1998,Nojiri2004,Zet2006}.  

Besides the theoretical dealing, in observations and experiments, studies are being pursued to understand the dynamics and morphologies of galaxies, galaxy formation and evolution, reionisation of the universe, history of the formation of the universe, the star formation, supernovae as cosmological probes, the formation of complex molecules in the diversity of astrophysical sources, simulation of the dense interstellar medium and so on. The parametrized post-Newtonian (PPN) formalism~\cite{Hohmann2014,Avilez2015,Viraj2017} is a tool, which furnishes a mathematical framework to solve the deviations of the experiments from GR and to what accuracy is verified by present-day observation. However, the PPN formalism is confined to only metric theories of gravitation. The measurement of the magnitude-redshift of distant supernovae is also a revolutionary observation~\cite{Kowalski2008,Amanullah2010,Suzuki2012}. From the data analysis of supernovae, it is clear that the decelerating parameter ($q$) lies in the range $-1.0 \leq q \leq -0.5$ of present days~\cite{Cunha2009,Li2011}. This result shows the reverse behavior of a universe described by GR where the negative result related to $q$ suggests the accelerating phase of the universe.

All these naturally suggest that to defend the observed data, it is necessary to flourish the GR theory in general aspects. Introduction of the extra dimensions or torsion or both of them to the spacetime is likely geometric extensions in GR. Nowadays, these theories are more relevant in the quantum field. In GR, the parallel transition of a system along the congruence of the observer is an evolution in time. In Finsler geometry, it is not necessary to be along the supporting direction. Finsler structure is the generalized form of the geometry which is not constrained to be linear in the second derivatives or quadratic in first derivatives of the metric, i.e. no quadratic restriction on measuring this infinitesimal distance~\cite{Chern1996}. 

The key idea for Finsler geometry for the measurement of time between two events which passes an observer is given by the length that connecting the events along observer's world-line. The measurement is based on the tangent bundle of a homogeneous function~\cite{Bao2000,Pfeifer2014}.  A few symmetric Finsler distance functions, instead of being defined on the tangent space by a Euclidean norm, is defined by a Minkowski p-norm. Apart from curvature, another property (color) of the manifold elaborates the geometrodynamics, which brings the intrinsic local anisotropy in addition to a few positional and directional dependent quantities~\cite{Kouretsis2012}. It is also important to distinguish particular properties from general ones.

Based on arc element, Riemann introduced a metric structure in a general space as
\[x 
 ds = F(x_1,~ : ~:~ x_n; dx_1,~ : ~:~ dx_n),
\]
where $F(x; y)$ is a positive function and homogeneous
of degree one in $y$. However, the important thing arises when 
\[
F^2=g_{\mu \nu} dx^i dx^j,
\]
which is the better description of Riemannian geometry without the quadratic restriction~\cite{Reddy2018}. Finslerian geometry extrapolates metric geometry (including Lorentz metric) by defining a length for the curve. Here, a general length measure for curves on $ \mathpzc{M}$ derives the geometry, instead of the metric. The whole concept was proposed by Finler in his thesis~\cite{Finsler1918}.

Several authors have considered Finsler geometry as a possible space-time model to define the relativistic behaviour. Vacuum field equation and the respective Schwarzschild solution is provided by Li~\cite{Li2014}. Silagadze~\cite{Silagadze2011} extends the Schwarzschild metric in Finsler space based on heuristic argument, which asymptotically approaches Bogoslovsky locally anisotropic space-time. Finslerian space-time is also suggested as the possible alternative to dark matter.  In the proposed modified Friedmann model, Chang et al.~\cite{Chang2009} showed that the accelerated expanding universe without invoking dark energy, and also provided modified newton's gravity in the frame of Finsler space with weak field approximation~\cite{Chang2008}. Javaloyes et al.~\cite{Javaloyes2014} investigated on the link between the geometry of spacetime and Finsler geometry. Usually, as in Brans-Dicke theory or string theory, the gravitational and physical geometry are conformally related Riemannian geometries. In this aspect Bekenstein~\cite{Bekenstein1993} explored that in weak equivalence principle and causality, the Finsler  geometry reduces to a  Riemann geometry whose physical metric, is related to the gravitational metric by a generalization of the conformal transformation involving a scalar field. 

In Finsler manifolds, the geometry of a concrete class and the conformal structure of a class of spacetimes are closely related. The geodesic motion of particles, following timelike and null geodesics as well as redshift relation on Finsler spacetimes with cosmological symmetry studied by Hohmann~\cite{Hohmann2017}. In low energy enclosure of supergravity theories, with $N$-connection structure, locally anisotropic spinors has been derived by Vacaru et al.~\cite{Vacaru1996,Vacaru1997,Vacaru2001,Vacaru2012}. 

To define the thermodynamical equilibrium state, as well as an equilibrium state Finsler geometry may be a possible solution. Finslerian curvature tensors are connected with the Cartan connection \cite{Mrugala1992} in the equilibrium state. In his pioneer work, Girelli~\cite{Girelli2007} shows that the Finsler geometry normally arises from quantum gravity ideas. Later on, Gibbons testifies this in special relativity~\cite{Gibbons2007}. Some authors are also using Finsler geometry to define the wormholes~\cite{Jusufi2017,Jusufi2018}. Cosmologically symmetric Finsler spacetime is discussed in details in the following articles~\cite{Pfeifer2011,Pfeifer2012,Hohmann2013,Hohmann2016}. 

On the other hand, singularity theorem and Raychaudhuri equation in the Finsler space-time is available in the work by Minguzzi~\cite{Minguzzi2015} whereas the generalized scalar-tensor theories in this spacetime are introduced by Stavrinos and Alexious~\cite{Stavrinos2018}. Recent observations of absorption spectra of quasar reveal that the fine structure constant ($\alpha = e^2/ \hslash c$) varies in the high redshift neighbourhood~\cite{Webb2011,King2012}. Mariano and Perivolaropoulos have shown the dipole alignment of fine structure constant with the dark energy dipole~\cite{Marino2012}. To describe the dipole structure of the fine structure constant Finsler geometry is a good natural theoretical account. 

Finsler modification of GR and the observable effects in a class of spherically symmetric and static spacetimes in the solar system also been reviewed by L{\"a}mmerzahl et al.~\cite{Hasse2012}. The PPN formalism for Finsler gravity was set up by Roxburgh~\cite{Roxburgh1992}. Aringazin and Asanov~\cite{Aringazin1985,Asanov1992} have generalized the  Schwarzschild metric in Finsler geometry, as well as studied the possible observational effects.

These are the huge motivation for considering Finsler geometry as a possible alternative spacetime model in different branches of science, especially to study the  issues of astrophysics, cosmology, GR and different gravities. Though, a complete analysis of the impact of a Finslerian modification of the geometry of spacetime on astrophysical as well as study of compact star especially strange star is still missing.
In this paper, we specifically would like to investigate the characteristics and behaviour of the strange star in the framework of Finsler geometry. The form of field equations is obtained by considering the kind of the flag curvature and the structure of the space. This geometry is totally engaged with the matter dynamics of the system and reduces to a Riemannian one if the metric tensor is assumed to be independent of dynamics. The construction of the field equations is interrelated with the internal variables, such as the velocity, spinors and different extra terms. The directional dependency of the curvatures on a scalar field is the outcome of the local anisotropy. As a consequence the pressure coefficient differs from the radial coordinate ($p_r(r)$) to the angular coordinates ($p_\theta(r)$, $p_\phi(r) \equiv  p_t(r)$). In this article, we also restricted ourselves to the MIT bag model equation of state (EOS) to define the strange stellar system. This is the simplest form of EOS to study stellar object made of up, down and strange quarks without conjuring any quantum mechanical aspect. The exterior of the stellar system is delineated by the Schwarzschild solution. Here, the respective radii are predicted from observable masses of the strange stars.  

The scheme of the study is as follows: We introduce the definition of Finsler geometry in Sec. II, and have provided the formalism of basic stellar equations in Finsler geometry. On the basis of this formalism, we have presented the solution to the Einstein field equations in Sec. III. Physical acceptability and stability of the stellar system are verified in Sec. IV by studying the mass-radius relation, energy conditions, stability of the stellar model and compactification factor. Finally, the conclusion of our study with a discussion is provided in Sec. V.

\section{BASIC STELLAR EQUATIONS}
We briefly present fundamental geometrical concepts from the theory of Finsler spaces and generate the respective field equations, as well as discuss about the EOS of the stellar system.

\subsection{Background Formalism}
Let us consider that $F$ is a Finsler metric on a manifold $ \mathpzc{M}$ which is defined as $F = F(x, \dot{x})$ and is a function of $(x^\mu, y^\mu)$ in  $\in T \mathpzc{M}$, in a standard coordinate system.

The Geodesic of the Finsler metric ($F$) is characterized by 
\begin{equation}
\frac{d^2x^\mu}{d \tau^2}+2G^\mu(x, \dot{x}) =0, \label{1}
\end{equation}
where the geodesic spray is 
\[G^\mu = \frac{1}{4} g^{\mu \nu} \left( \frac{\partial ^2F^2}{\partial x^\lambda \partial y^\nu} y^\lambda -\frac{\partial F^2}{\partial x^\nu} \right).
\] 

The metric structure coefficient can be written in the form
 \begin{equation}
 g_{\mu \nu} = \frac{\partial }{\partial y^\mu} \frac{\partial }{ \partial y^\nu} \left( \frac{1}{2} F^2 \right), \label{2}
 \end{equation}
where $(g ^{\mu \nu} )$ = $ (g_{\mu \nu})^{-1} $ and also note that each $ g_{\mu \nu} $ is homogeneous of degree zero in $y$. 
 
For a non-zero vector $ y=y^{\mu}(\frac{\partial}{\partial x^\mu})\mid_{p} \in T_{p} \mathpzc{M}, ~F $ induces an inner product on $ T_{p} \mathpzc{M} $ which is given by
 \[
g_{y}(u,v)=g_{\mu \nu}(x,y)u^\mu v^\nu,
 \]
where $ u=u^\mu(\frac{\partial}{\partial x^\mu})\mid_{p},~ v=v^\mu (\frac{\partial}{\partial x^\mu})\mid_{p}  \in T_{p} \mathpzc{M} \setminus \lbrace 0 \rbrace$. 

Let the Finsler structure is of the form
\begin{equation}
F^2= e^{\lambda(r)} y^t y^t - e^{\nu(r)} y^r y^r -r^2 \overline{F}^2(\theta, \phi, y^\theta, y^\phi ). \label{3}
\end{equation}

Therefore, with the help of Eq. (\ref{1}), the Finsler metric potential can be written as
 \begin{eqnarray}
&g_{\mu \nu} = diag(e^{\lambda(r)},~- e^{\nu(r)}, ~-r^2\overline{g}_{ij}). \label{4} 
\end{eqnarray}

Respective geodesics sprays are as follows
\begin{eqnarray}
& G^t = \frac{1}{2}\lambda' y^t y^r, \label{5} \\
& G^r = \frac{1}{4}  \left( \nu' y^r y^r +\lambda' e^{\lambda-\nu} y^t y^t -2re^{-\nu}\overline{F}^2  \right),  \label{6} \\
& G^\theta = \frac{1}{r} y^\theta y^r +\overline{G}^\theta, \label{7}\\
& G^\phi = \frac{1}{r} y^\theta y^r +\overline{G}^\phi.  \label{8}
\end{eqnarray}

Ricci scalar of the Finsler geometry can be written as
{\footnotesize 
\begin{equation}
Ric =R^\mu_\mu=\frac{1}{F^2} \left[ 2 \frac{\partial G^\mu}{\partial x^\mu}- y^\lambda \frac{\partial^2 G^\mu}{\partial x^\lambda \partial y^\mu} +2 G^\lambda \frac{\partial^2 G^\mu}{\partial y^\lambda \partial y^\mu} - \frac{\partial G^\mu}{\partial y^\lambda}\frac{\partial G^\lambda}{\partial y^\mu} \right], \label{9}
\end{equation}} 
where $R^\mu_\mu$ is insensitive to connections which only depends on Finsler structure.

Ricci tensor of Finsler geometry is introduced by Akbar-Zadeh~\cite{Akbar1988} as follows
\begin{equation}
Ric_{\mu\nu}= \frac{\partial^2}{\partial y^\mu \partial y^\nu}\left( \frac{1}{2}F^2 Ric \right). \label{10}
\end{equation}

On  considering the values of $G^\mu$ from  Eqs. (\ref{5})-(\ref{8}) and substituting in Eq. (\ref{9}), we obtain
\begin{widetext}
{\small{
\begin{eqnarray}
F^2 Ric= \left[ \frac{\lambda''}{2}+\frac{\lambda'^2}{4}-\frac{\lambda' \nu'}{4} +\frac{\lambda'}{r} \right] e^{(\lambda-\nu)} y^t y^t + \left[- \frac{\lambda''}{2}-\frac{\lambda'^2}{4} +\frac{\lambda' \nu'}{4} +\frac{\nu'}{r} \right]y^r y^r +(\overline{Ric}-e^{-\nu} +\frac{r \nu' e^{-\nu}}{2}-\frac{r \lambda' e^{-\nu}}{2}). \label{11}
\end{eqnarray}
}}
\end{widetext}

On the other hand, the scalar curvature is defined as $S = g^{\mu \nu} Ric_{\mu \nu}$.

Therefore, the Einstein tensor in Finsler spacetime takes the modified form
\begin{eqnarray}
G^{\mu }_{\nu} = g^{\mu \nu}Ric_{\mu \nu}-\frac{1}{2} S, \label{12}
\end{eqnarray}
with the choice of the flag curvature in the form  $ \overline{F}^2 =y^\theta y^\theta + f(\theta) y^\phi y^\phi $ (vide $Appendix~B$).

Hence, in the explicit form we have
\begin{multline}
S = e^{-\nu}\left[ \lambda''+\frac{\lambda'^2}{2}- \frac{\lambda' \nu'}{2} +\frac{\lambda'-\nu'}{r} \right] \\ -\frac{2}{r^2} \left[\overline{Ric}-e^{-\nu} +\frac{r  e^{-\nu}}{2} (\nu'- \lambda')\right]. \label{13}
\end{multline}

The energy-momentum tensor for the anisotropic fluid distribution of strange quark matter (SQM) can be written in the form
\begin{equation}
T^{\mu}_{\nu}= (\rho +p_t)u^{\mu}u_{\nu} - p_t \delta^{\mu}_{\nu} - (p_t-p_r)v^{\mu}v_{\nu}, \label{14}
\end{equation}
where $\rho$, $p_r$ and $p_t$ represent the energy density, radial and tangential pressures respectively for the fluid system whereas            $ u_{\nu} $ and $ v_{\nu} $ represent the four-velocity and radial four-vector, respectively.

The covariant divergence of the stress-energy tensor is
\begin{equation}
 T^\mu_{~\nu;\mu} = 0. \nonumber
\end{equation}

Following the notion of GR, the field equations in the Finsler space-time geometry can be provided as
\[
G^\mu_{\nu}  = 8 \pi_F T^\mu_{\nu}.
\]

With the help of Eqs. (\ref{5})-(\ref{8}) and (\ref{12})-(\ref{14}), the Einstein field equations for an anisotropic stellar system are in the form
\begin{align} \centering
	 \frac{\nu' e^{-\nu}}{r} - \frac{ e^{-\nu}}{r^2} +\frac{\overline{Ric}}{r^2} &= 8 \pi_F \rho, \label{15a}\\
 \frac{\lambda' e^{-\nu}}{r} + \frac{ e^{-\nu}}{r^2} - \frac{\overline{Ric}}{r^2} &= 8 \pi_F p_r, \label{16a}\\
 e^{-\nu}\left[ \frac{\lambda''}{2}+\frac{\lambda'^2}{4}-\frac{\lambda' \nu'}{4} +\frac{\lambda'-\nu'}{2r} \right] &= 8 \pi_F p_t, \label{17a} \\
  e^{-\nu}\left[ \frac{\lambda''}{2}+\frac{\lambda'^2}{4}-\frac{\lambda' \nu'}{4} +\frac{\lambda'-\nu'}{2r} \right] &= 8 \pi_F p_t. \label{18a} 
\end{align}

For simplicity, we consider the density profile of the fluid inside the strange stars maintaining the form defined by Mak and Harko~\cite{Mak2002} as follows
\begin{equation}
\rho(r)=\rho_c\left[1-\left(1-\frac{\rho_0}{\rho_c}\right)\frac{r^{2}}{R^{2}}\right],\label{19}
\end{equation}
where $\rho_c$ and $\rho_0$ are the central and surface densities respectively.

In a general way, $m(r)$ is the mass-energy contained within the Finser sphere of radius $r$, at the selected symmetry of time and is provided in the form~\cite{Li2014}
\begin{equation}
m(r) = \int_{0}^{r} 4\pi_Fr^2\rho(r) dr. \label{20a}
\end{equation}

\subsection{Junction Condition}
The stellar structure equations can be integrated from the centre towards the surface of the system. The constrains of the system are as follows
 \[
   m(0) = 0,    ~~ \rho(0) = \rho_{c} ~~   and    ~~ p_r(0)=p_c.
 \]

The surface (${\it r = R}$) of the system is determined by $ p_r(R)=0 $. Now, the interior spacetime should be matched smoothly with the exterior vacuum space time along the junction, i.e. surface of the defined system. The Schwarzschild  metric to represent the exterior spacetime in the Finslerian geometry can be written in the following form~\cite{Li2014}
{
\begin{multline}
F^2= \left(1-\frac{2GM}{CR}\right) y^t y^t - \left( C-\frac{2GM}{R} \right) ^{-1} y^r y^r \\-r^2 \overline{F}^2(\theta, \phi, y^\theta, y^\phi ) \label{20},
\end{multline}
}
where $M$ is the total mass of the system and $C$ is a constant (with the geometrisized unit $G=1$).

\subsection{Equation of state (EOS)}
We consider that the matter distribution inside the strange stars is regularized by phenomenological MIT bag model EOS as proposed by Chodos et al.~\cite{Chodos1974}. The three flavoured quarks in the bag are considered as non-interacting and massless.

Therefore, the total quark pressure can be defined as
\[
p_r = {\sum_f}{p^f} - {B_g},
\]
where $ p^f $ is the pressure of the up ($u$), down ($d$) and strange ($s$) quarks respectively and $B$ is the vacuum energy density. Here, $ p^f =\frac{1}{3}{{\rho}^f}$ is the pressure of individual quarks whereas $ {{\rho}^f}$ is the energy density of the individual quarks. 

Again, the energy density of individual de-confined quarks is as follows
\[
{\sum_f}{{\rho}^f}=\rho+B_g.
\]

Hence, from the above two equations, eventually we get
\begin{equation}
p_r=\frac{1}{3}(\rho-4B_g). \label{21}
\end{equation}

The co-relation of energy and pressure of SQM is maintained by {\it ad hoc} bag function.

Since, the radial pressure must vanish on the surface, therefore from Eq. (\ref{21}) we can conclude
\[
\rho_0 = 4B_g
\]
where $ \rho_0 $ is the surface density (i.e. at $r= R$).

Hence, we obtain the modified form as follows
\begin{equation}
p_r=\frac{1}{3}(\rho-\rho_0). \label{22}
\end{equation}

\section{SOLUTION TO THE FIELD EQUATIONS}
From Eqs. (\ref{14})-(\ref{19}) and with the help of EOS of strange star Eq. (\ref{21}), we obtain
\begin{widetext}
\begin{equation}
	\nu(r) = - \ln \left[ \overline{Ric}-\frac{8}{3} \pi {r}^{2}\rho_{{c}}+\frac{8}{5 {R}^{2}} {\pi {r}^{4} \left( \rho_{{c}}- 4 B_g \right) } \right], \label{23}
\end{equation} 

	{\footnotesize
\begin{eqnarray}
		\lambda(r) =  \frac{64}{3 \lambda_1} \Bigg\{\left(- \frac{1}{32} \ln  \left(  \left( 16 B_g\pi {r}^{2}+\overline{Ric} \right) {R}^{5}-16 B_g\pi {R}^{3}{r}^{4}-5M{R}^{2}{r}^{2}+3M{r}^{4} \right) + {\frac {3}{64}\ln  \left( \overline{Ric}-2 {\frac {M}{R}} \right) } +  \frac{1}{32} \ln \left( \overline{Ric}{R}^{5}-2 M{R}^{4} \right)  \right) \lambda_1 \nonumber \\  + \frac{1}{32} {R}^{2} \left( 32 B_g\pi {R}^{3}-5M \right)\left( \arctanh \left({\frac {16B_g\pi {R}^{5}-32B_g\pi {R}^{3}{r}^{2}-5M{R}^{2}+6M{r}^{2}}{\lambda_1}}\right)  - \arctanh \left( {\frac {-16B_g\pi {R}^{5}+M{R}^{2}}{\lambda_1}}	\right)  \right) \Bigg\}, \label{24}
\end{eqnarray} 
	}
	
\begin{equation}
p_r =\frac{1}{3} \left( \rho_{{c}}-\rho_{{0}} \right)  \left( 1-{\frac {{r}^{2}}{{R}^{2}}} \right), \label{25}
\end{equation}
	
\begin{eqnarray}
p_{t}=\frac{1}{45}{\frac {16 \pi {C_{1}}^{2}{r}^{6}-32 \left( \rho_{{0}}
			-\frac{5}{4}\rho_{{c}} \right) \pi {R}^{2}C_{1} {r}^{4}+30 {R}^{2}
			\left( C_{2} {R}^{2}+\overline{Ric} C_{1} \right) {r}^{2}-15 \overline{Ric} {R}^{4}C_{1}}{
			\left( -\frac{8}{5} \pi C_{1} {r}^{4}-\frac{8}{3} \pi {R}^{2}{r}^{2}\rho_{{c}}+\overline{Ric} 
			{R}^{2} \right) {R}^{2}}}, \label{26}
\end{eqnarray}
	
\begin{equation}
\Delta=\frac{1}{3} \frac{\left[  \left( \frac{2}{3} \left( \rho_{{0}}-2 \rho_{{c}} \right)  \left( 
		-4 \rho_{{c}}+\rho_{{0}} \right) {R}^{4} - \frac{56}{15}{C_{1}  {r}^{2}{R}^{2} \left( \rho_{{0}}-{
				\frac {13}{7}} \rho_{{c}} \right) }+\frac{8}{3} {r}^{4} C_{1}^{2} \right) \pi+\overline{Ric} {R}^{2} C_{1} \right] {r}^{2} }{{R}^{2} \left( -\frac{8}{3} \left( \rho_{{c}}{R}^{2}+\frac{3}{5} C_{1} {r}^{2} \right) {r}^{2}\pi+\overline{Ric} {R}^{2} 
			\right) } \label{27}
\end{equation}	
\end{widetext}
where $ \lambda_1$, $ C_1 $ and $ C_2  $ are the constants which are provided in the $Appendix~B$.

\begin{figure}
	\includegraphics[scale=0.4]{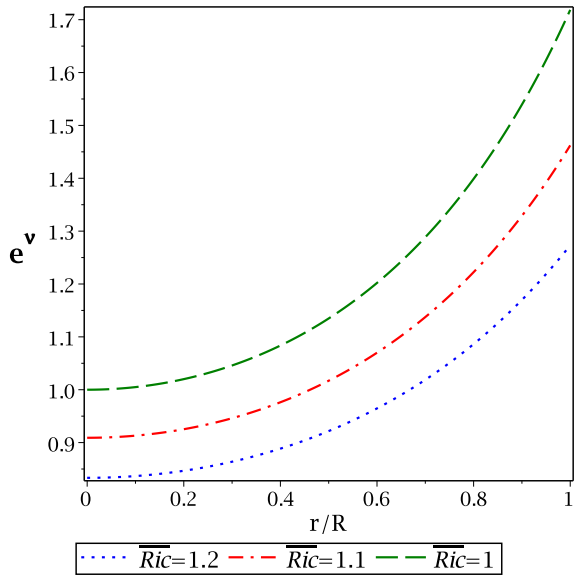}
	\caption{Variation of $e^{\nu(r)}$ as a function of the fractional radial coordinate $r/R$, with bag constant (B$_g$) = 100~MeV/fm$^3$ for the $LMC~X-4$.} \label{fn}
\end{figure}

\begin{figure}
	\includegraphics[scale=0.4]{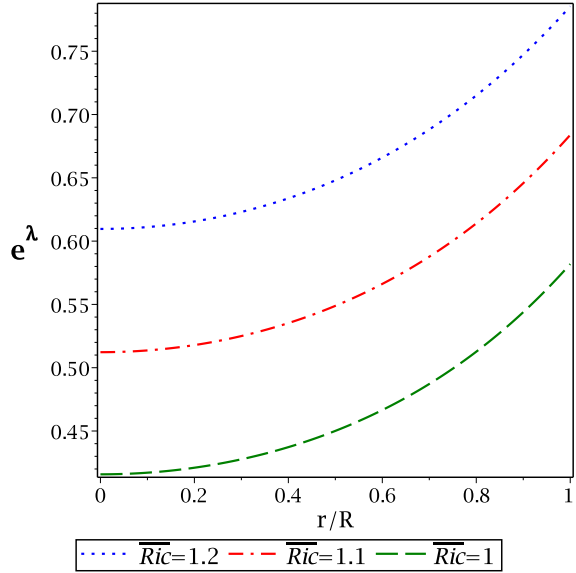}
	\caption{Variation of $e^{\lambda(r)}$ as a function of the fractional radial coordinate $r/R$, with bag constant (B$_g$) = 100~MeV/fm$^3$ for the $LMC~X-4$.} \label{fl}
\end{figure}

The variations of the gravitational potentials of the following space-time geometry are exhibited in Figs. \ref{fn} and \ref{fl} with respect to the fractional radial coordinate in connection to the respective Eqs. (\ref{23}) and (\ref{24}).

\begin{figure}
	\includegraphics[scale=0.4]{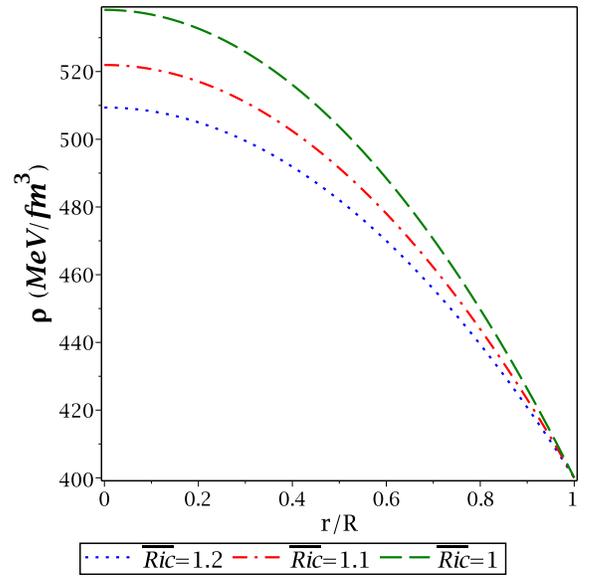}
	\caption{Variation of the density $(\rho)$ as a function of the fractional radial coordinate $r/R$, with bag constant (B$_g$) = 100~MeV/fm$^3$ for the $LMC~X-4$.} \label{frho}
\end{figure}

\begin{figure}
	\includegraphics[scale=0.4]{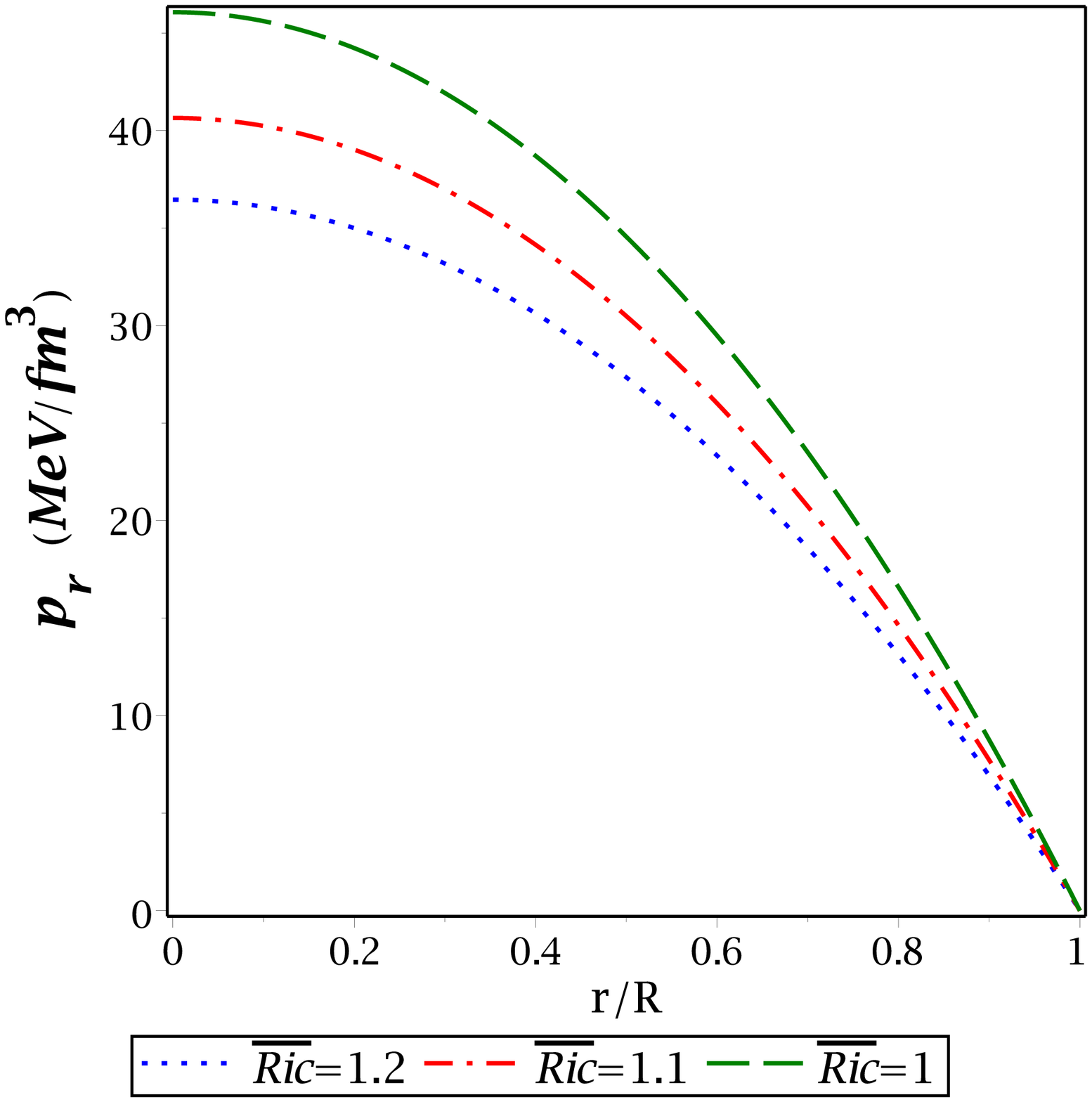}
	\caption{Variation of the radial pressure $(p_r)$ as a function of the fractional radial coordinate $r/R$, with bag constant (B$_g$) = 100~MeV/fm$^3$ for the $LMC~X-4$.} \label{fpr}
\end{figure}

\begin{figure}
	\includegraphics[scale=0.4]{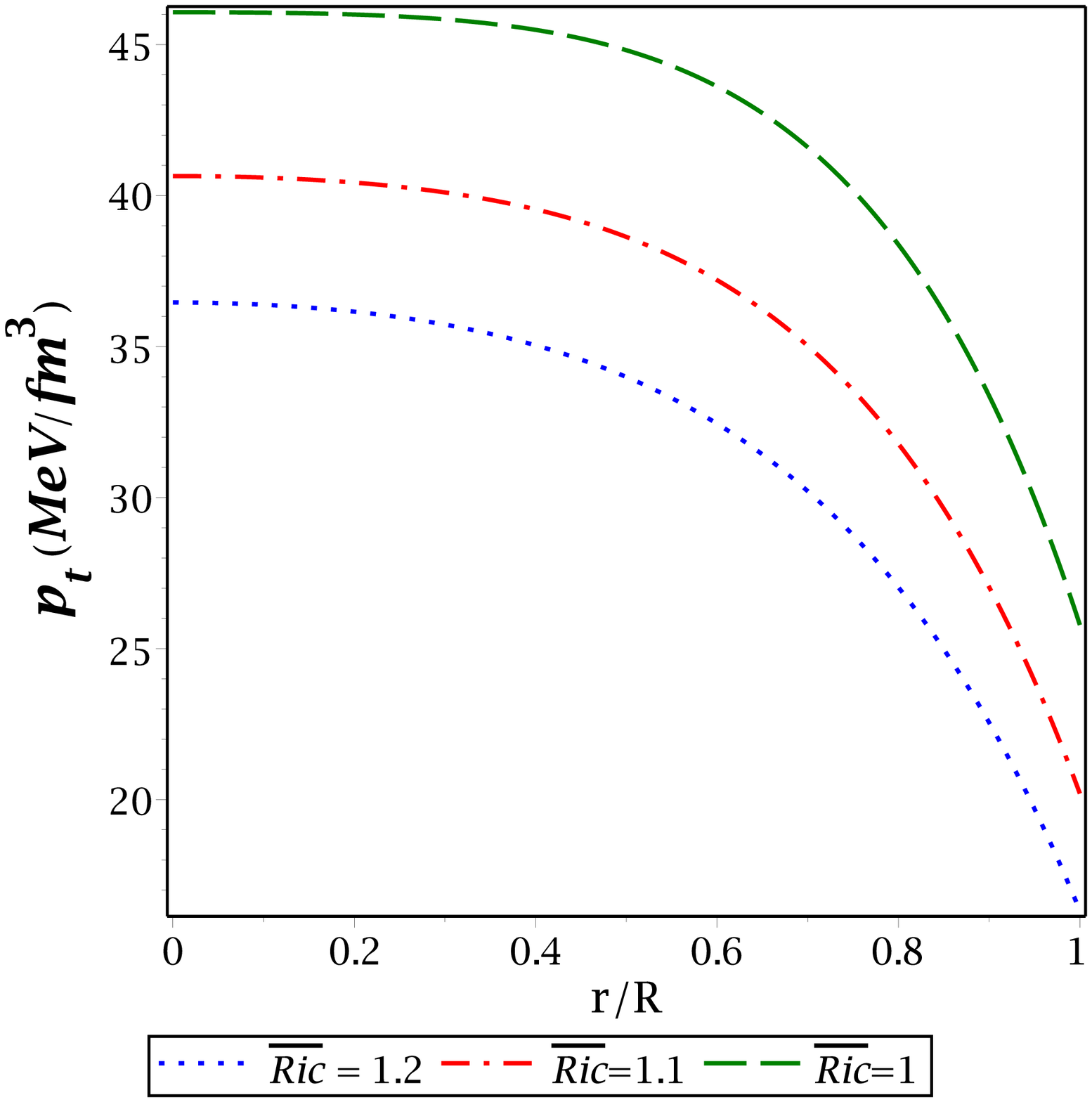}
	\caption{Variation of the tangential pressure $(p_t)$ as a function of the fractional radial coordinate $r/R$, with bag constant (B$_g$) = 100~MeV/fm$^3$ for the $LMC~X-4$.} \label{fpt}
\end{figure}

The variations of the physical quantities like radial pressure, tangential pressure and density are shown in Figs. \ref{fpr}, \ref{fpt} and \ref{frho} respectively in reference of fractional radial coordinate where the expressions are presented by Eqs. (\ref{25}), (\ref{26}) and (\ref{19}).

The anisotropy ($ \Delta $) of a system is defined as the excessive stress along the tangential direction over radial direction, i.e.
$\Delta(r)=p_t-p_r$. The explicit expression of the anisotropy is provided in Eq. \ref{27}. 

\begin{figure}
	\includegraphics[scale=0.4]{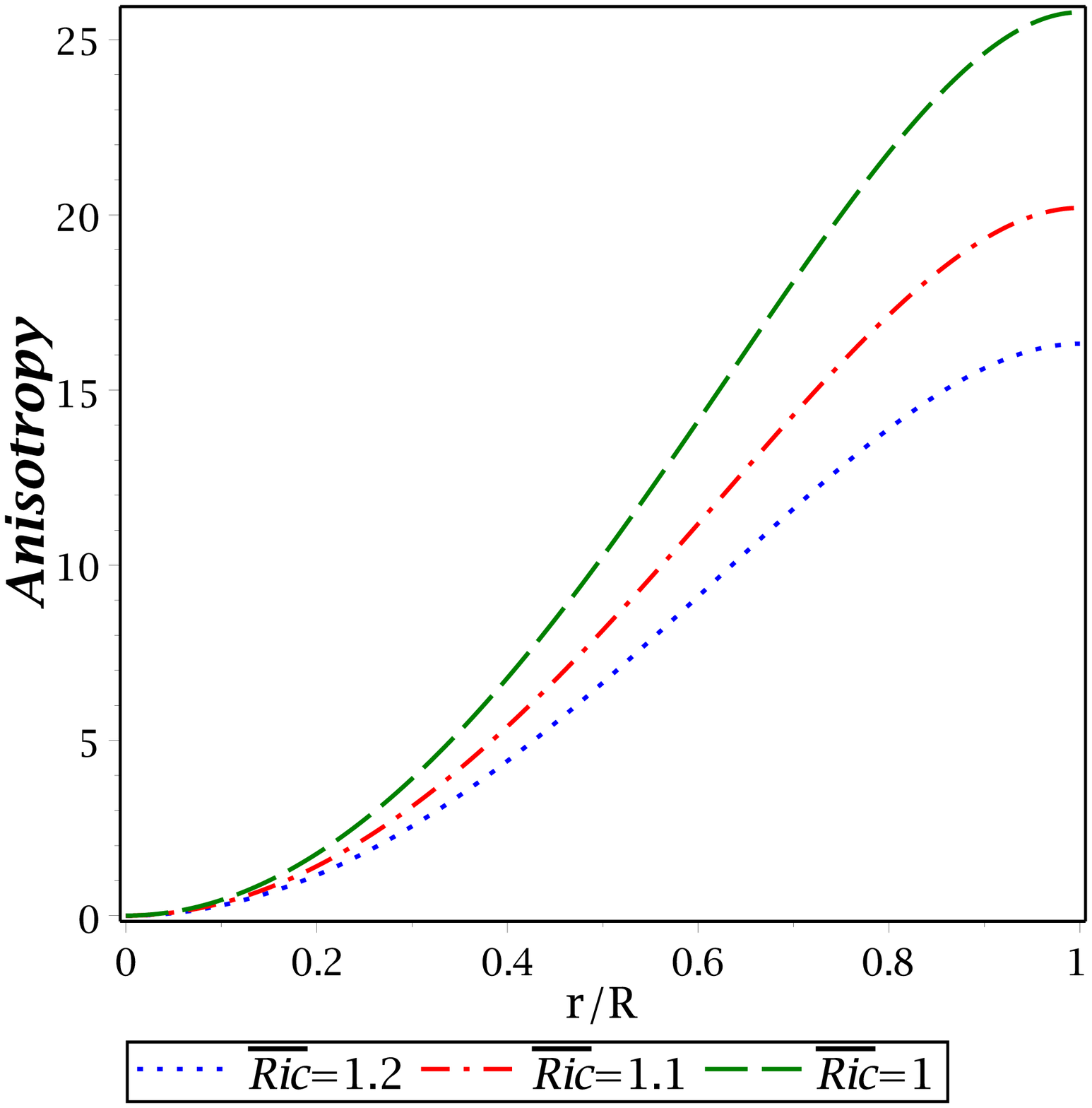}
	\caption{Variation of the anisotropic stress $(\Delta)$ as a function of the fractional radial coordinate $r/R$, with bag constant (B$_g$) = 100~MeV/fm$^3$ for the $LMC~X-4$.} \label{faniso}
\end{figure}

The anisotropy of the present stellar system is exhibited in Fig.~\ref{faniso}, which depict that the anisotropy is increasing monotonically at the center (from the minimum value) and achives its maximum at the surface. This is a well predicted result in GR.

\section{PHYSICAL BEHAVIOUR OF THE SYSTEM}

\subsection{Stability of the system}
To study the stability of the system we consider the following issues: (a) TOV equation, (b) Herrera cracking condition and (c) Adiabatic index.

\subsubsection{TOV equation}
A system is considered to be in equilibrium, if the summation of all active forces on the system is null, proposed by Tolman~\cite{Tolman1939} and Oppenheimer-Volkoff~\cite{OV1939}.

Therefore, according to TOV equation
\begin{equation}
-p_r'-\frac{\lambda'}{2}(\rho+p_r)+\frac{2}{r}(p_t-p_r) = 0, \nonumber
\end{equation}
where the first term defines the hydrostatic force ($  F_{h} $), the second term stands for gravitational force ($  F_{g} $) and the last term defines the anisotropic force ($  F_{a} $) of the system. Here, the outward forces $  F_{a} $ and $  F_{h} $ are balancing the inward pull $  F_{g} $.

\begin{figure}
	\includegraphics[scale=0.4]{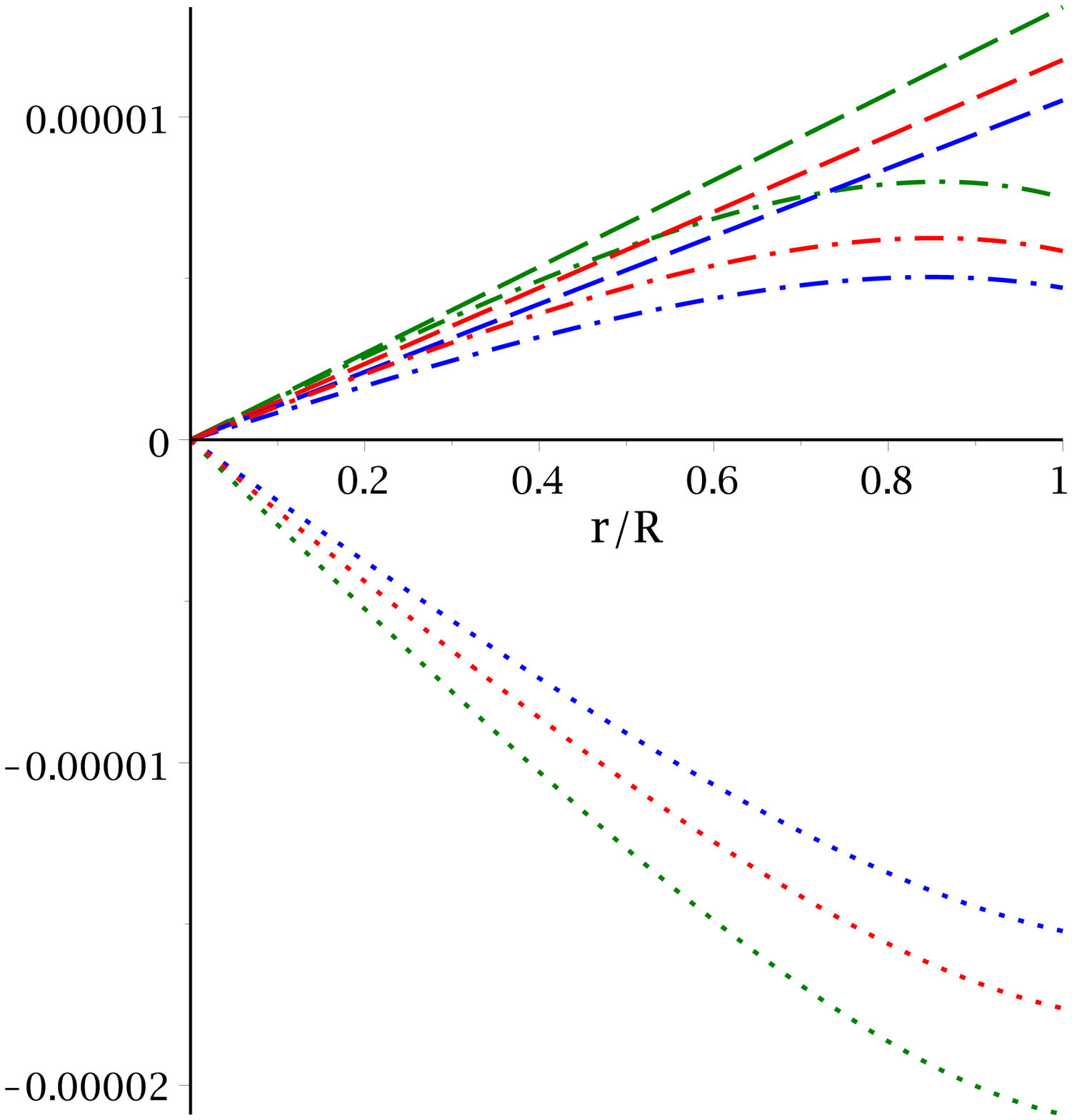}
	\includegraphics[scale=0.4]{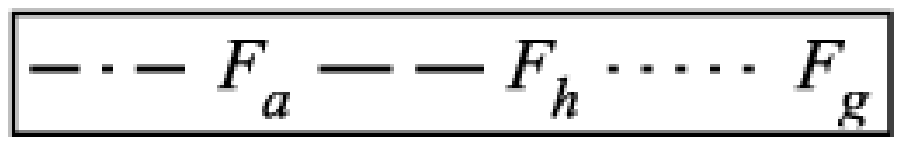}
	\caption{Variation of different forces $(F_i)$ as a function of the fractional radial coordinate $r/R$, with bag constant (B$_g$) = 100~MeV/fm$^3$ for the $LMC~X-4$. Here green: $\overline{Ric}$=1, red: $\overline{Ric}$=1.1, blue: $\overline{Ric}$=1.2.} \label{ff}
\end{figure}

The equilibrium of the forces is achieved for every values of $ \overline{Ric} $ which confirms the stability of the system. The variation of the forces is shown in Fig.~\ref{ff}.

\subsubsection{Herrera cracking condition}
For a physically stable system, the causality condition should be maintained. The condition states that the square of the radial  ($ v_{sr}^2 $) and the tangential ($ v_{st}^2 $) sound speed lies between 0 $\rightarrow$ 1, i.e. 0 $\leq v_{si}^2 \leq 1 $ (where $i=r, t$). For a potentially stable region the radial sound speed must be greater than the tangential sound speeds. According to Herrera~\cite{Herrera1992} and Abreu \cite{Abreu2007}  the difference of square of the sound speed should kept the sign inside the stellar system. Hence, according to Herrera’s condition $ \mid v_{st}^2 -v_{sr}^2 \mid \leq $ 1. 

\begin{figure}
	\includegraphics[scale=0.4]{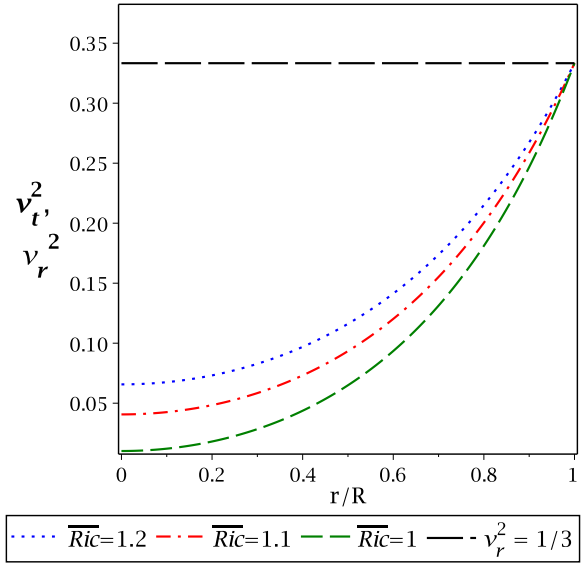}
	\caption{Variation of the sound velocities $(v_{si}^2)$ as a function of the fractional radial coordinate $r/R$, with bag constant (B$_g$) = 100~MeV/fm$^3$ for the $LMC~X-4$.} \label{fs}
\end{figure}

\begin{figure}
	\includegraphics[scale=0.4]{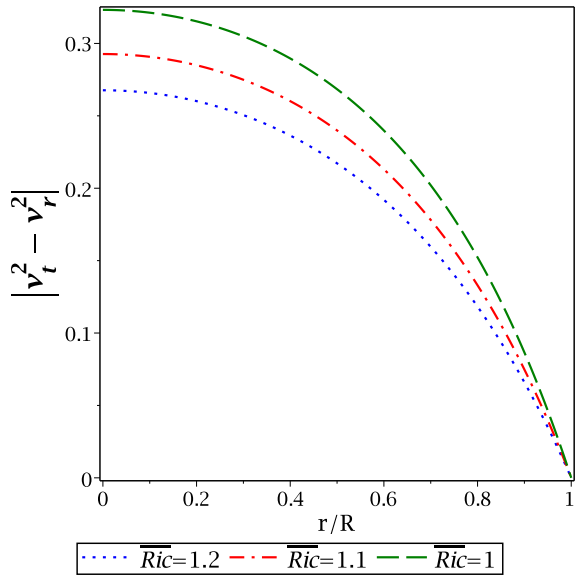}
	\caption{Variation of the metric coefficient $e^{\nu(r)}$ as a function of the fractional radial coordinate $r/R$, with bag constant (B$_g$) = 100~MeV/fm$^3$ for the $LMC~X-4$.} \label{fh}
\end{figure}

Variation of the radial and tangential sound speed and the respective differences are shown in Figs. \ref{fs} and \ref{fh}, respectively.

\subsubsection{Adiabatic index}
The stability of a star can also defined with the help of adiabatic index ($ \Gamma $) of the system. It characterized the stiffness of the EOS for a particular density variation. Following Chandrasekhar~\cite{Chandrasekhar1964}, several authers~\cite{Santos1997,Horvat2011,Doneva2012,Silva2015} also studied the dynamical stability for an infinitesimal radial perturbation. Heintzmann and Hillebrandt~\cite{Heintzmann1975} showed that for a stable stellar system the adiabatic index must be greater than $4/3$. 

Now, the adiabatic index is defined as 
 \begin{eqnarray}
 \Gamma = \left( \frac{p_r + \rho }{p_r}\right)  \frac{dp_r}{d \rho}  = \frac{p_r + \rho }{p_r} v_{sr}^2, \nonumber
 \end{eqnarray}
 
\begin{figure}
	\includegraphics[scale=0.4]{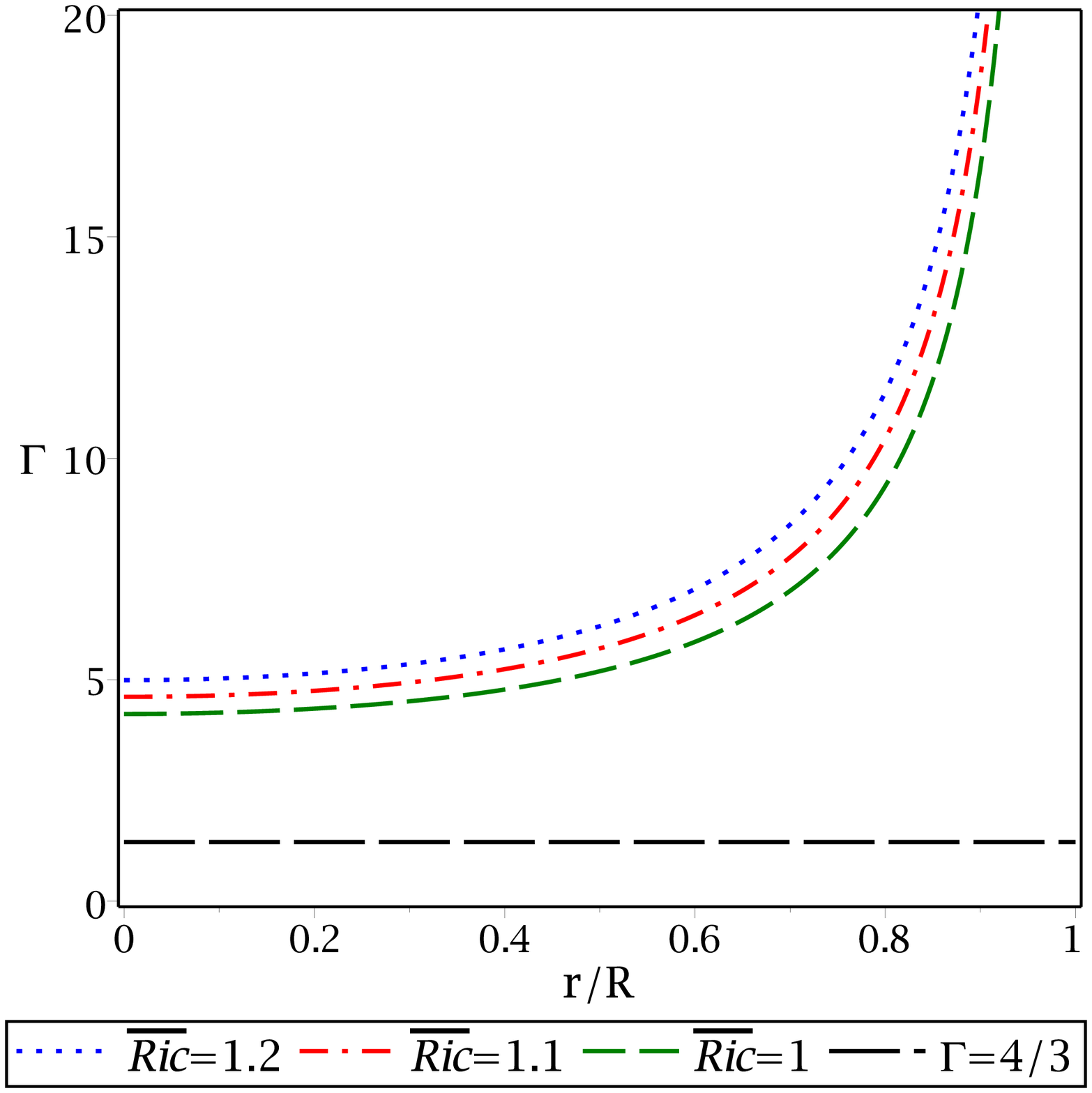}
	\caption{Variation of adiabatic index as a function of the fractional radial coordinate $r/R$, with bag constant(B$_g$) = 100MeV/fm$^3$ for the $LMC ~X-4$.} \label{fa}
\end{figure}

Variation of the adiabatic index with the fractional radial coordinate is shown in Fig.~\ref{fa}

\subsection{Energy conditions}
The system will be a physically valid system, if it satisfied the following inequalities: 
\begin{align}
\qquad\hspace{-0.5cm}~NEC&:\rho+p_r\geq 0,~\rho+p_t\geq 0, \nonumber \\ 
\qquad\hspace{-0.5cm}~WEC&: \rho+p_r\geq 0,~\rho\geq 0,~\rho+p_t\geq 0, \nonumber \\
\qquad\hspace{-0.5cm}~SEC&: \rho+p_r\geq 0,~\rho+p_r+2\,p_t\geq 0, \nonumber \\ \qquad\hspace{-0.5cm}~DEC&:\rho\geq 0,~ {{\rho}-{p_r}}\geq 0, ~{{\rho}-{p_t}}\geq 0. \nonumber
\end{align}

Here NEC (null energy condition), WEC (weak energy condition), SEC (strong energy condition) and DEC (dominant energy condition).

The energy conditions for the Finslerian system are provided by Stavrinos and Alexiou~\cite{Stavrinos2018}.

\begin{figure}
	\includegraphics[scale=0.4]{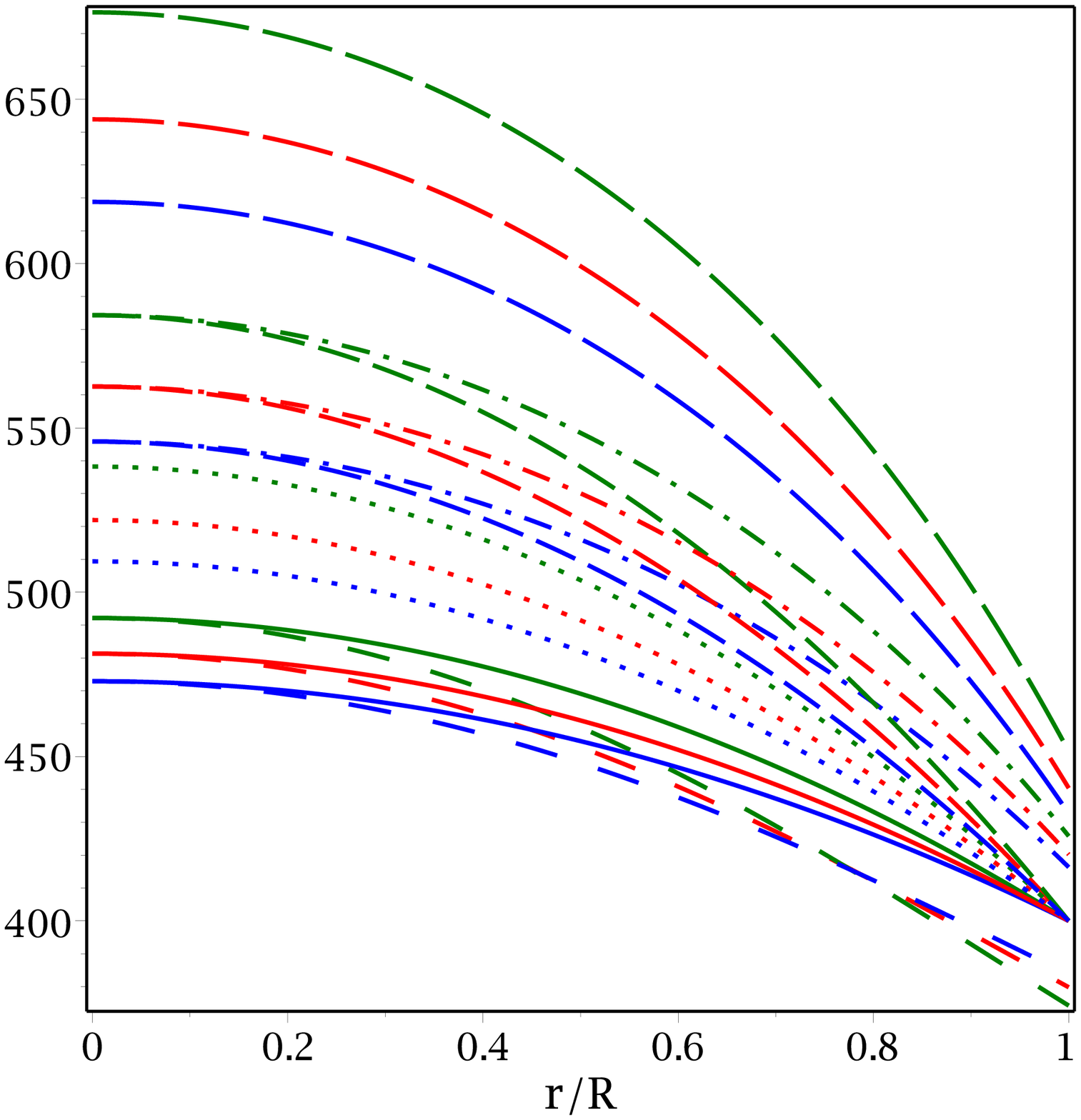}
	\includegraphics[scale=0.3]{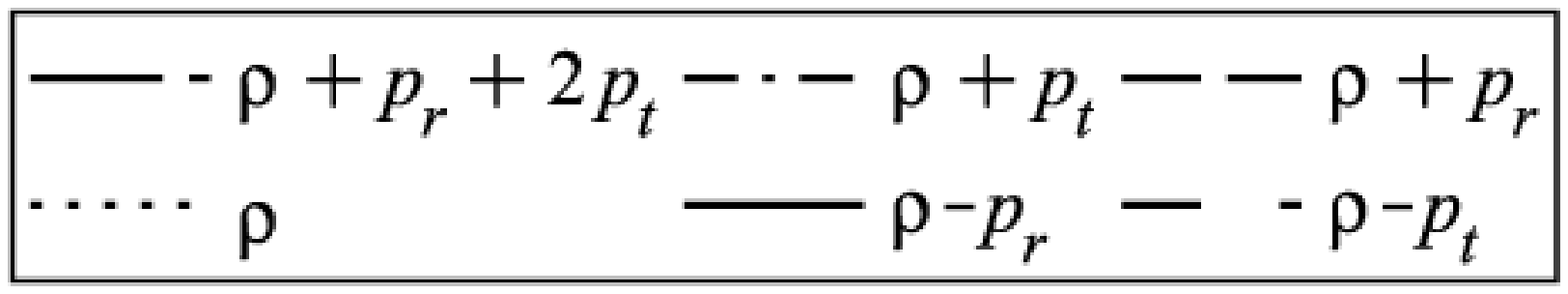}
	\caption{Variation of different energy as a function of the fractional radial coordinate $r/R$, with bag constant (B$_g$) = 100~MeV/fm$^3$ for the $LMC ~X-4$. Here green: $\overline{Ric}$=1, red: $\overline{Ric}$=1.1, blue: $\overline{Ric}$=1.2.} \label{fe}
\end{figure}

The variation of the different energy conditions with the fractional radial coordinate is shown in Fig. \ref{fe}.

\subsection{Fractional Binding Energy}
The gravitational binding energy of a stellar system can defined as the minimum amount of energy required for the system to cease being in a gravitationally bound state. It depends on the mass as well as radius of the system. In case of stellar system the density distribution is always non-uniform. The fractional gravitational binding energy for the non-uniform distribution is defined by Lattimer and Prakash~\cite{Lattimar2001} as 
 \begin{equation}
  f_b=  \frac{0.6 \beta}{1- 0.5 \beta}, \label{28}
 \end{equation}
where $ \beta $ is defined as $M/R$. 

\begin{figure}
	\includegraphics[scale=0.4]{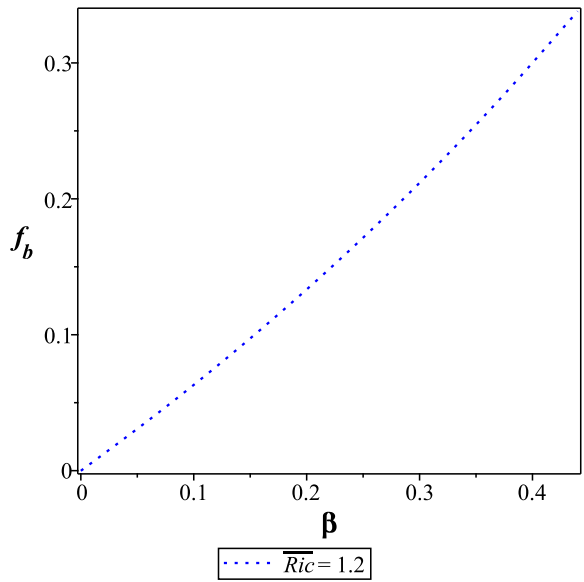}
	\caption{Variation of the fractional binding energy$(f_b)$ as a function of  $m/r$, with bag constant (B$_g$) = 100~MeV/fm$^3$.} \label{fb}
\end{figure}

Variation of $f_b$ as a function of the total gravitational mass of a strange star for is displayed in Fig. \ref{fb}.

\subsection{Redshift}
The compactification factor is defined by $ u(r)=m/r $ and expressed as
\[ 
u = \frac {1}{r} \left( \frac{4}{3}\pi {r}^{3}\rho_{{c}}-\frac{4}{5}{\frac {\pi {
			r}^{5} \left( \rho_{{c}}-\rho_{{0}} \right) }{{R}^{2}}} \right).
 \]

\begin{figure}
	\includegraphics[scale=0.4]{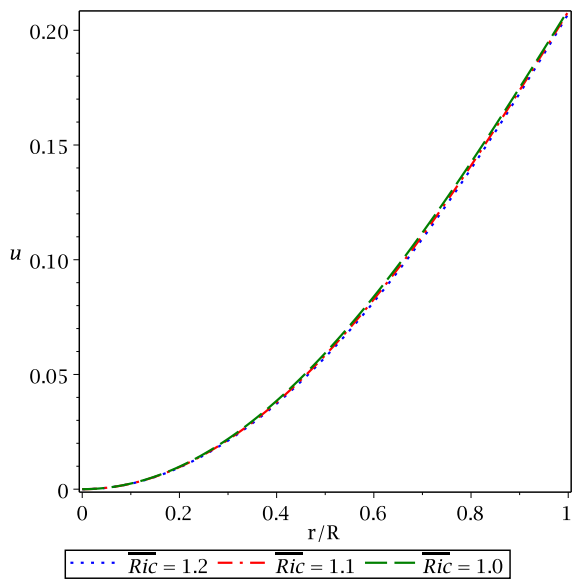}
	\caption{Variation of compactification factor $(u)$ as a function of the fractional radial coordinate $r/R$, with bag constant (B$_g$) = 100~MeV/fm$^3$ for the $LMC ~X-4$} \label{fc}
\end{figure}

Variation of the compactness is shown in Fig. \ref{fc}

The expression of the gravitational redshift of the system is in the form
\begin{widetext}
Z= e$^{-\lambda /2}$-1
 {\footnotesize
\begin{align}
= \exp \Bigg[-\frac{32}{3 \lambda_1} \Bigg\{\left(- \frac{1}{32} \ln  \left(  \left( 16 B_g\pi {r}^{2}+ \overline{Ric} \right) {R}^{5}-16 B_g\pi {R}^{3}{r}^{4}-5 M{R}^{2}{r}^{2}+3 M{r}^{4} \right) +  \frac{1}{32} \ln \left( \overline{Ric} {R}^{5}-2 M{R}^{4} \right) + {\frac {3}{64}\ln  \left( \overline{Ric}-2 {\frac {M}{R}} \right) } \right) \lambda_1 \nonumber \\  + \frac{1}{32} {R}^{2} \left( 32\,B_g\pi  {R}^{3}-5 M \right)\left( \arctanh \left({\frac {16 B_g\pi  {R}^{5}-32 B_g\pi {R}^{3}{r}^{2}-5 M{R}^{2}+6 M{r}^{2}}{\lambda_1}}\right)  - \arctanh \left( {\frac {-16 B_g\pi {R}^{5}+M{R}^{2}}{\lambda_1}}	\right)  \right) \Bigg\} \Bigg]-1.
\end{align} 
 }
\end{widetext}

\begin{figure}
	\includegraphics[scale=0.4]{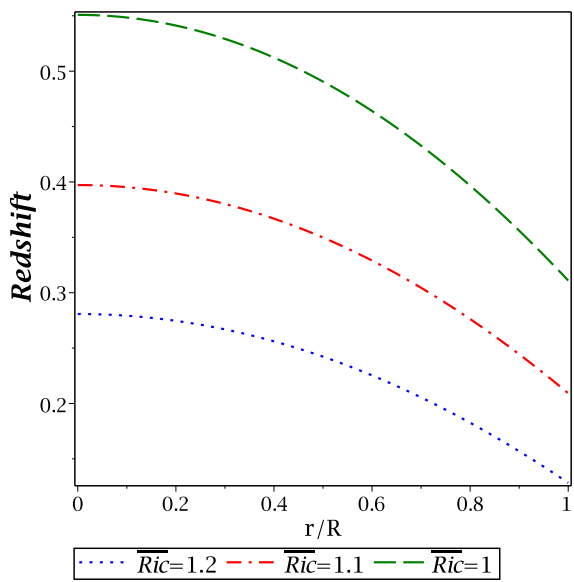}
	\caption{Variation of redshift $(Z_s)$ as a function of the fractional radial coordinate $r/R$, with bag constant (B$_g$) = 100~MeV/fm$^3$ for the $LMC ~X-4$} \label{fr}
\end{figure}

Variation of the redshift with the fractional radial coordinate is shown in Fig. \ref{fr}.

\subsection{Mass-Radius relationship }
For any stable star, in all regions $2m(r)/r \leq 1$ should be maintained~\cite{Hartle1973}. The maximum bounded mass of a perfect fluid system is proposed by Buchdahl~\cite{Buchdahl1959} as $2M/R \leq 8/9$. Later, Mak and Harko~\cite{Mak2003} generalized the expression.

\begin{figure}
	\includegraphics[scale=0.4]{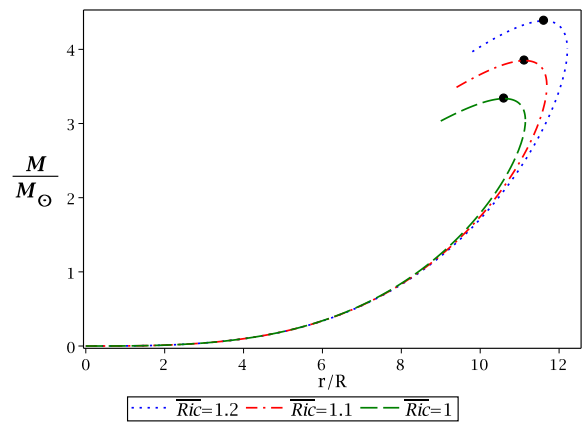}
	\caption{Variation of the mass of a strange star as a function of radius. Solid circles are representing the maximum mass and radius of the respective curves. Here curves are drawn for $B_g$ = 100~MeV/fm$^3$} \label{fmr}
\end{figure}

We have shown variation of the total mass ($M$), normalized in solar masses (M$_\odot $), with respect to the total radius ($R$) in Fig. \ref{fmr} for different values of $ \overline{Ric} $ for bag constant $ B_{g} $ = 100 MeV/fm$ ^{3} $. The variation achieves same as in GR. It is clear from the graph that the maximum mass and respective radius increases gradually with increasing value of $ \overline{Ric} $.

\section{DISCUSSION AND CONCLUSION}
 In the present article we find the strange star in account of the geometry which is unbounded to quadratic in first derivatives or linear in the second derivatives of the metric, on measuring the infinitesimal distance. 
 
The metric potentials, viz., $ e^{\nu(r)} $ and $ e^{\lambda(r)} $ featured in Figs. \ref{fn} and \ref{fl} indicate that our stellar system is geometrically non-singular system and the potentials are monotonically increasing functions. The density, radial component of pressure and tangential component of pressure are monotonically decreasing inside of the stellar system from maximum values at the centre, shown in the Figs. \ref{frho}, \ref{fpr} and \ref{fpt}, respectively. No overlying matter against the gravitational attractive force applied onto it from the bulk inside endorse no radial pressure and this marks as the edge of the system. From the expression, provided in Eq. (\ref{27}) and the plot \ref{faniso},  we acquire that the anisotropic stress reaches maximum at the surface from the minimum value (zero) at the centre of the system as predicted in GR by Deb et al.~\cite{Deb2017}. The physical acceptability of the system is verified on the basis of TOV equation, cracking condition, fractional binding energy, energy conditions, mass-radius relation. 
 
 The net force acting outward is the anisotropic flow in addition with the pressure gradient is supported by the gravity acting on the shell by the matter (mass) inwards to it. It defines the hydrostatic equilibrium of the fluid element which is at rest within the system, as well as the overlying matter which decreases with radial coordinate. The variation of all the forces are featured in Fig. \ref{ff}, which validates that our system is unvarying in terms of the equilibrium of forces.

 The term cracking  basically refers to the proper way of deviation from equilibrium. The notion itself is independent of the geometry and the general conditions of its occurrence also, though depends on the specific theory of gravitation. State of equilibrium is constrained by the condition that the difference must lie from zero to one. Throughout the stellar system the following conditions are to be maintained. We found that our system is consistent with (i) the causality relation and (ii) the Herrera cracking concept. The variation is portrayed in Figs. \ref{fs} and \ref{fh}, respectively. 

 Gravitational redshift is related with the metric potential of the system. It is monotonically decreasing with the radial variation towards the surface. Variation of (i) compactification factor and (ii) redshift as a function of the radial coordinate $r = R$ for the strange star $LMC~X-4$ are shown in Figs. \ref{fc} and \ref{fr}, respectively.

 It is useful to express the fractional gravitational binding energy ($f_b$) as a function of the compactness
 parameter of a stellar system. It contributes to the mass of the star significantly, thus radius varies with $f_b$. This is also strongly intralinked with the underlying EOS. 

 Energy conditions state the energy density of the matter distribution as measured by an observer in the space-time and they must be positive. This represent the notion that the matter should flow along null or time-like world line. In this system all energy conditions are preserved conveniently. 

 The total mass $M$ (normalized in M$ _\odot $) of a stellar system with respective to the corresponding radius ($R$) for chosen values of $ \overline{Ric} $ and $ B_g=100MeV/fm{^3} $ is presented in Fig. \ref{fmr}. The solid circles in the plot represent the maximum mass points of the corresponding radius for the strange stars. From Fig. \ref{fmr} it is clear that the mass and radius gradually increases with the increasing values of $ \overline{Ric} $. We find for $ \overline{Ric} $ = 1, the maximum mass is 3.27 M$ _\odot $ and the respective radius is 10.53 km. However, it is to be noted that for $ \overline{Ric} $ = 1.2, the mass is increased by 24.35\% and the respective radius is increased by 10.02\%. In each of these variations the mass-radius ratio is less than the singularity condition. 
  
 In Table \ref{tab:table1}, we have predicted the radius and different physical properties to explain the stars with observed mass. The Table is developed with the constrains B$ _g $ = 100 Mev/fm${^3}$ and $ \overline{Ric} $ = 1.2. From the surface redshift and compatification factor of the stars, it is clear that they are highly dense in nature.

\begin{table*}
 	\caption{\label{tab:table1} The physical parameters are predicted from the proposed model where $1~{{M}_{\odot}}=1.475~km$ for $G=c=1$. The set is valid for B$ _g $ = 100 Mev/fm${^3}$ and $ \overline{Ric} $ = 1.2.}
 	\begin{ruledtabular}
 		\begin{tabular}{cccccccc}
 			Strange&Observed&Predicted&${\rho}_{c}$&${{P_c}}$&Fractional binding&Red-shift  \\
 			Stars&Mass$({{M}_{\odot}})$&Radius (km)&(${gm/cm}^3$)&(${dyne/cm}^2$)&Energy($f_b$)&  ($Z_s$) \\ \hline \hline \\
 			
 			$PSR~J~1614-2230$ & $1.97 \pm 0.04$\cite{Demorest2010} & $10.42^{+0.06}_{-0.07}$  & $9.95 \times {10}^{14} $ & $ 8.49 \times {10}^{34} $ &  0.194 &  0.248 \\ 
 			
 			$Cyg~X-2$ & $1.78 \pm 0.23$\cite{Orosz1999} &  $10.13 ^{+0.36}_{-0.41}$ & $9.69 \times 10^{14}$ & $7.72 \times 10^{34}$ & 0.179 & 0.211 \\
 			
 			$PSR~B1913+16$ & $1.44 \pm 0.002$\cite{Leeuwen2015} &  $9.49 ^{+0.01}_{-0.01}$ & $9.25 \times 10^{14}$ & $6.39 \times 10^{34}$ & 0.151 & 0.153 \\
 			
 			$Cen~X-3$ & $1.09 \pm 0.08$\cite{Kerkwijk1995} &  $8.72 ^{+0.19}_{-0.21}$ & $8.82 \times 10^{14}$ & $5.11 \times 10^{34}$ &  0.122 & 0.097 \\ 
 			
 			$LMC~X-4$ & $1.29 \pm 0.05$\cite{Gangopadhyay2013} &  $9.18 ^{+0.11}_{-0.12}$ &  $9.09 \times 10^{14}$ & $5.91\times 10^{34}$ & 0.139 & 0.128 \\ 
 			
 			$4U~1820-30$ & $1.58 \pm 0.06$\cite{Guver2010a} &  $9.76 ^{+0.11}_{-0.11}$ &  $9.42 \times 10^{14}$ & $6.92 \times 10^{34}$ & 0.163 & 0.185 \\
 			
 			$4U~1608-52$ & $1.74 \pm 0.14$\cite{Guver2010b} &  $10.05 ^{+0.23}_{-0.25}$ & $9.64 \times 10^{14}$ & $7.55 \times 10^{34}$ & 0.176 & 0.206 \\ 
 			
 			$EXO~1785-248$ & $1.30 \pm 0.20$\cite{Ozel2009} &  $9.20 ^{+0.41}_{-0.46}$ & $9.07 \times 10^{14}$ & $5.87 \times 10^{34}$ & 0.140 & 0.130 \\
 			
 			$Vela~X-1$  & $1.77 \pm 0.16$\cite{Barziv2001} &  $10.10^{+0.26}_{-0.28}$  & $ 9.67 \times 10^{14} $ & $7.67 \times 10^{34}$ & 0.178 & 0.210 \\ 
 		\end{tabular}
 	\end{ruledtabular}
 \end{table*}

 \begin{figure}
	\includegraphics[scale=0.4]{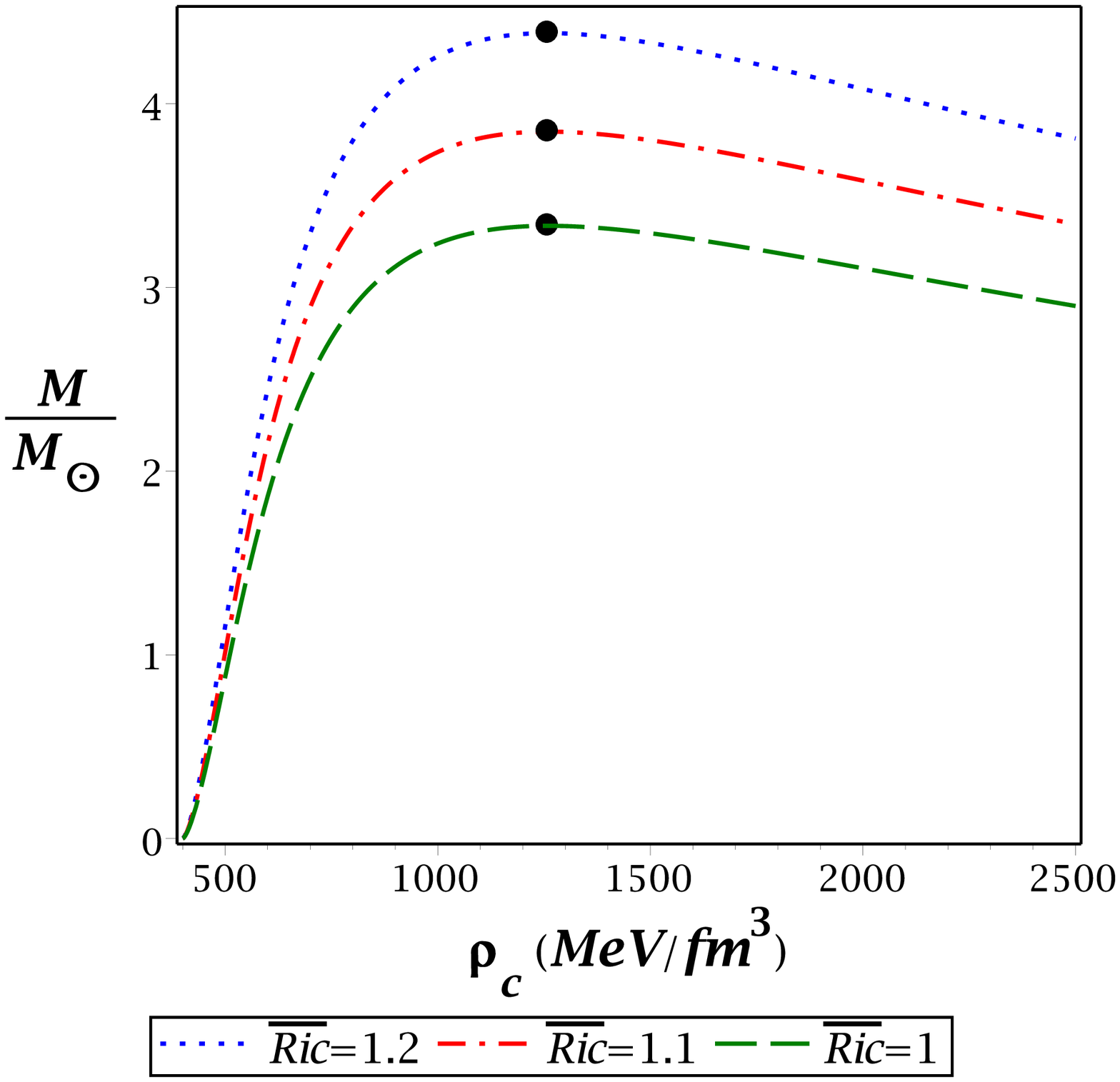} 
	\caption{The variation of mass of a strange star as a function of the central density $({\rho}_c)$. Solid circles are representing the maximum mass and radius of the respective curves. Curves are drawn for $B_g$ = 100~MeV/fm$^3$}  \label{fmrho}
\end{figure}

\begin{figure}
	\includegraphics[scale=0.4]{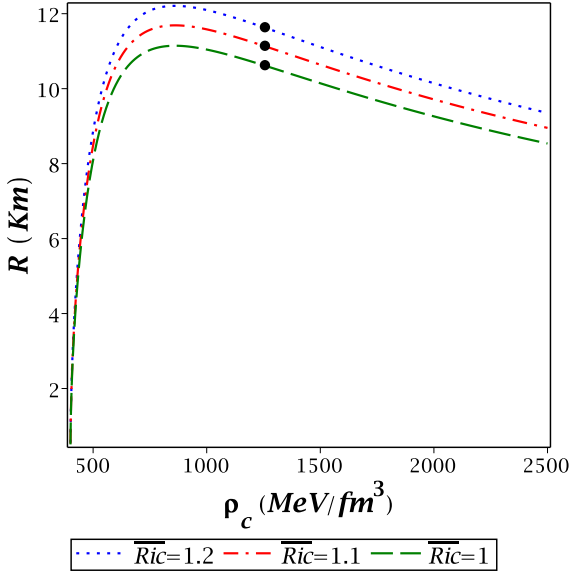}
	\caption{The variation of radius of a strange star as a function of the central density $({\rho}_c)$. Solid circles are representing the maximum mass and radius of the respective curves. Curves are drawn for $B_g$ = 100~MeV/fm$^3$}  \label{frrho}
\end{figure}

  Above behaviour states that the equilibrium conditions are satisfied by the stellar system. Although, equilibrium is not the assurance of stability, it corresponds to either maximum or minimum in energy with respect to radial variation.  An equilibrium configuration can be a stable configuration depending on a few conditions. 

 An equilibrium configuration should satisfy the microscopic stability condition of a matter, i.e. dp/d$\rho  > $ 0. The stability of an astronomical objects are examined against the radial oscillation (study of adiabatic index of the star). The adiabatic index ($ \Gamma $) = d(ln p$ _{r} $)/d(ln $ \rho $) does express the equilibrium configuration as well as stability of the star. Variation of the adiabatic index $\Gamma$ with respect to the radial coordinate $r/R$ shown in Fig. \ref{fa}. From this it can be concluded that the adiabatic index is greater than 4/3, so our system is stable against the radial pulsation. 

 For a specific value of the central density, a stellar system can move from stable equilibrium configuration to unstable equilibrium configuration at which the equilibrium mass is stationary. The necessary condition for stability, the stellar system should lie in the region $\frac{dM}{d \rho_{{c}}} > 0$. Fig. \ref{fmrho} shows the stellar mass in M$_\odot$ alongside the central density for different values of $\overline{Ric}$. Full dots over the curves signify the point where the maximum mass values are found. We seize the central densities in the range 400 $ \leq \rho_{{c}} \leq $ 2500 MeV/fm$ ^{3} $. From plot we can enunciate that the mass of the stellar system increases with the increment of the central energy density until it reaches the uttermost value which is 1258.29 MeV/fm$ ^{3} $ with 3.27 M$_\odot$ for $\overline{Ric}$ = 1. It is clear from the graph that our stellar system satisfies the stability condition. With increasing B$_g$ the maximum central density increases, though the mass decreases, as already predicted in GR. The variation of radius against the central densities for different $\overline{Ric}$ is shown in Fig. \ref{frrho}. With the increasing $\overline{Ric}$, the maximum value of the radius is also raising.

 \begin{table*}
 	\centering \caption{\label{tab:table2} Numerical values of the physical parameters for different {$\overline{Ric}$} for the strange star $LMC~X-4$ of mass 1.29$M_\odot$ ($1~{{M}_{\odot}}=1.475~km$) with B$ _g $ = 100 Mev/fm${^3}$.}
 	\begin{ruledtabular}	
 		\begin{tabular}{c  c  c  c  c  c  c}
 			
 			Value of $\overline{Ric}$  &\hspace{0.55cm} $ \overline{Ric} $ = 1 & \hspace{0.55cm} $ \overline{Ric} $ = 1.1 & \hspace{.55cm} $ \overline{Ric} $ = 1.2 \\ \hline \hline\\  
 			
 			Predicted Radius($km$) &\hspace{0.55cm} 9.10 &\hspace{0.55cm} 9.15 &\hspace{0.55cm}  9.18  \\
 			
 			$\rho_c~({gm/cm}^3)$ &\hspace{0.55cm} $9.57\times 10^{14} $ &\hspace{0.55cm}  $9.29\times 10^{14} $ &\hspace{0.55cm} $9.09\times 10^{14} $ \\ 
 			
 			$P_c~({dyne/cm}^2)$ &\hspace{0.55cm} $7.37\times 10^{34} $ &\hspace{0.55cm} $6.51\times 10^{34} $ &\hspace{0.55cm} $5.91\times 10^{34} $ \\ 
 			
 			$\frac{2M}{R}$  &\hspace{0.55cm} 0.418 &\hspace{0.55cm} 0.416 &\hspace{0.55cm} 0.414 \\ 
 			
 			Fractional Binding Energy($f_b$) &\hspace{0.55cm} 0.140 &\hspace{0.55cm} 0.139 &\hspace{0.55cm} 0.138 \\
 			
 			Red Shift($Z_s$) &\hspace{0.55cm} 0.311 &\hspace{0.55cm} 0.209 &\hspace{0.55cm} 0.128 \\ 
 			
 		\end{tabular}
 	\end{ruledtabular}
 \end{table*}

We constraint ourselves in the flag curvature. The defined flag curvature (the quantity involves in a certain location $x \in  \mathpzc{M}$, a flag pole ${\ell } := y/F(y)$ with $y \in T_p \mathpzc{M}$ and a transverse edge $V \in T_p \mathpzc{M}$) is independent of the actual length of edge of the flag pole~\cite{Bao2000}. There is no loss of generality in choosing of this. We have considered that $ \overline{Ric} $ as a constant quantity following Li and Chang~\cite{Li2014}. The usefulness of constant flag curvature is explained by Akbar-Zahed~ \cite{Akbar1988}. The reason of choosing the value of $ \overline{Ric}  \geq 1$ is that the system is not providing the physically acceptable solution below the critical value. 

In a general way, $m(r)$ is the mass-energy contained with in the Finser sphere of radius $r$, at the selected symmetry of time. It is provided in the form $m(r) = \int_{0}^{r} 4\pi_Fr^2\rho(r) dr $~\cite{Li2014}.
 
In Table \ref{tab:table2} we predict different physical parameters for the observed mass of $LMC~X-4$ for B$ _g $ = 100 Mev/fm${^3}$ under the chosen values of $ \overline{Ric} $ as 1, 1.1 and 1.2. We find that as the parameter $ \overline{Ric} $ increases the mass (M) and respective radius (R) of the star gradually. However, Table \ref{tab:table2} shows that with increasing value of $ \overline{Ric} $ the central density ($ \rho_c $), central pressure (p$_{c} $), surface redshift ($ Z_s $), fractional binding energy ($ f_b $) and the value of $2M/R$ decreases gradually.

The variation of the radius as well as central density, central pressure, fractional binding energy, redshift and compactness for $LMC~X-4$ due to different Bag values with  constant $ \overline{Ric} $ (= 1.2) is shown in Table \ref{tab:table3}. It is clear from the table that the parameters are varying with B{$_g$}'s, for higher Bag values the system become more compressive (higher compactification factor as well as fractional binding energy) and as a result the radius of the star decreases.

\begin{table*}
	\centering \caption{\label{tab:table3} Numerical values of the physical parameters for different Bag values for the strange star $LMC~X-4$ of mass 1.29$M_\odot$ ($1~{{M}_{\odot}}=1.475~km$) with $ \overline{Ric} $ = 1.2.}
	\begin{ruledtabular}	
		\begin{tabular}{c  c  c  c  c  c  c}
			
			Value of B$ _g $  &\hspace{0.55cm} B$ _g $ = 90 Mev/fm$ ^3 $ & \hspace{0.55cm} B$ _g $ = 100 Mev/fm$ ^3 $ & \hspace{.55cm} B$ _g $ = 110 Mev/fm$ ^3 $ \\ \hline \hline\\  
			
			Predicted Radius($km$) &\hspace{0.55cm} 9.52 &\hspace{0.55cm} 9.18 &\hspace{0.55cm}  8.88  \\
			
			$\rho_c~({gm/cm}^3)$ &\hspace{0.55cm} $8.06\times 10^{14} $ &\hspace{0.55cm}  $9.09\times 10^{14} $  &\hspace{0.55cm} $10.02\times 10^{14} $ \\ 
			
			$P_c~({dyne/cm}^2)$ &\hspace{0.55cm} $5.02\times 10^{34} $ &\hspace{0.55cm} $5.91\times 10^{34} $ &\hspace{0.55cm} $6.65\times 10^{34} $ \\ 
			
			$\frac{2M}{R}$  &\hspace{0.55cm} 0.400 &\hspace{0.55cm} 0.415 &\hspace{0.55cm} 0.429 \\ 
			
			Fractional Binding Energy($f_b$) &\hspace{0.55cm} 0.133 &\hspace{0.55cm} 0.138 &\hspace{0.55cm} 0.144 \\
			
			Red Shift($Z_s$) &\hspace{0.55cm} 0.119 &\hspace{0.55cm} 0.128 &\hspace{0.55cm} 0.139 \\ 
			
		\end{tabular}
	\end{ruledtabular}
\end{table*}

The redshift has an inverse relation with the parameter $ \overline{Ric}$. It is clear from the tables that the high B$ _g $ as well as high $ \overline{Ric} $ produce a more compact strange star. We also find that a stellar system becomes more massive in Finsler framework, compared to Riemannian system. Also finite variation of the central density and central pressure are established.

In the domain of weak gravity, Einstein's theory is well verified from laboratory-based experiments and solar system tests. Though, in case of strong gravity, viz. in the vicinity of a black hole, expanding universe, ultra dense compact stars etc., the theory still has constraints. In the recent scenario, the intensely luminous supernovas, e.g. $SN~2003fg$, $SN~2006gz$, $SN~2007if$ and $SN~2009dc$~\cite{Howell2006,Scalzo2010} which have been discovered as a part of $SNLS$, hint a huge $Ni$-mass and support the super-Chandrasekhar white dwarfs of mass 2.1-2.8 M$_\odot$~\cite{Hicken2007,Yamanaka2009,Silverman2011,Taubenberger2011}. In observation of binary $PSR~J2215+5135$, Linares et al.~\cite{Linares2018} discovered a massive pulsar of 2.27$^{+0.17}_{-0.15}$ M$_\odot$. These observations indicate the requirement of modification of Einstein's theory in strong gravity. In the Finslerian frame, the expansion of the universe and its anisotropic behaviour has been studied~\cite{Kouretsis2009,Kouretsis2010,Li2015}. Interestingly, our study reveals that in the Finsler frame, the maximal mass can be higher than their standard values in Riemannian frame for the chosen parametric values of $\overline{Ric}$. Hence, the stellar systems in the Finslerian geometry may explain the observed massive stellar system.

Finally, we can enunciate that in this paper we have successfully presented a stable and physically acceptable anisotropic stellar model in the framework of Finsler geometry which seems suitable to strange stars. It is worth mentioning that MIT bag model takes a definite role to address the strange star candidates in alternative geometry.

\appendix

\section{Flag Curvature}
Let us choose, the flag curvature in the form
	\[
	 \overline{F}^2 =y^\theta y^\theta + f(\theta) y^\phi y^\phi .\label{a}
	\]
	
	\begin{eqnarray}	
	&\overline{g}_{ij} = (1,f(\theta)), \nonumber \\ 
	&\overline{g}^{ij} = (1,f(\theta)^{-1}). \nonumber 
	\end{eqnarray}
	
Now, considering $(\dot{~})= \frac{\partial}{\partial \theta}$, we define
	\begin{eqnarray}
	\overline{G}^\theta = -\frac{1}{4} \dot{f}(\theta)y^\phi y^\phi,  \label{b}
	\end{eqnarray}
	
	\begin{eqnarray}
	\overline{G}^\phi = \frac{1}{2} \frac{\dot{f}(\theta)}{f(\theta)} y^\theta y^\phi,  \label{c}
	\end{eqnarray}
		
	\begin{equation}
	\frac{\partial \overline{G}^\mu}{\partial x^\mu} = -\frac{1}{4} \ddot{f}(\theta)y^\phi y^\phi, \label{d}
	\end{equation}
	
	\begin{equation}
	y^\lambda \frac{\partial^2 \overline{G}^\mu}{\partial x^\lambda \partial y^\mu} = \frac{1}{2} \left( \frac{\ddot{f}(\theta)}{f(\theta)}- \left(\frac{\dot{f}(\theta)}{f(\theta)}\right)^2 \right) y^\theta  y^\theta \label{e},
	\end{equation}
		
	\begin{equation}
	\overline{G}^\lambda \frac{\partial^2 \overline{G}^\mu}{\partial y^\lambda \partial y^\mu} = -  \frac{1}{8} \frac{\dot{f}(\theta)^2}{f(\theta)} y^\phi y^\phi  \label{f}, 
	\end{equation}
	
	\begin{equation}
	\frac{\partial \overline{G}^\lambda}{\partial y^\mu} \frac{\partial \overline{G}^\mu}{\partial y^\lambda} =   \frac{1}{4}\left(\frac{\dot{f}(\theta)}{f(\theta)}\right)^2  y^\theta  y^\theta -  \frac{1}{2} \frac{\dot{f}(\theta)^2}{f(\theta)} y^\phi y^\phi.  \label{g} 
	\end{equation}
	
With the help of Eqs. [(\ref{d})-(\ref{g})], we can write
	\begin{align}
	\overline{F}^2 \overline{Ric}= - \frac{1}{2} \ddot{f}(\theta)y^\phi y^\phi -\frac{1}{2} \left( \frac{\ddot{f}(\theta)}{f(\theta)}- \left(\frac{\dot{f}(\theta)}{f(\theta)}\right)^2 \right) y^\theta  y^\theta \nonumber \\ - \frac{1}{4} \frac{\dot{f}(\theta)^2}{f(\theta)} y^\phi y^\phi - \frac{1}{4}\left(\frac{\dot{f}(\theta)}{f(\theta)}\right)^2  y^\theta  y^\theta +  \frac{1}{2} \frac{\dot{f}(\theta)^2}{f(\theta)} y^\phi y^\phi,
	\end{align}

	\begin{equation}
	\overline{Ric}= \left[ \frac{1}{4}\left(\frac{\dot{f}(\theta)}{f(\theta)}\right)^2 -  \frac{1}{2} \frac{\ddot{f}(\theta)}{f(\theta)} \right]\label{h}.
	\end{equation}

\section{Expression of the constants}
\begin{eqnarray}
C_{1}=\rho_{{0}}-\rho_{{c}},
\end{eqnarray}

\begin{equation}
C_{2}=\frac{1}{3}\pi \left( {\rho_{{0}}}^{2}-2\rho_{{c}}\rho_{{0}}+4{
	\rho_{{c}}}^{2} \right),
\end{equation}

\begin{align}
\lambda_1= \Bigg[ 256{B_{g}}^{2}{\pi }^{2}{R}^{10}+64 \overline{Ric} B_{g}\pi {R}^{8}-160 B_{g}M\pi 
		{R}^{7}\nonumber \\-12\overline{Ric} M{R}^{5}+25 {M}^{2}{R}^{4}\Bigg] ^{1/2}, 
\end{align}
{               
 \begin{align}    
 \rho_{{c}} = \frac{1}{32}\frac{1}{ {R}^{2}\pi \left( 8\pi {R}^{2}B_{g}+\overline{Ric} \right)} \Bigg[ -256{\pi }^{2}{R}^{4}{B_{g}}^{2}\nonumber\\
		 	        +112\pi {R}^{2}B_{g}\overline{Ric} + 15 \overline{Ric}^{2}- \Bigg\{1638400{\pi }^{4}{R}^{8}B_{g}^{4}\nonumber \\-204800{\pi }^{3}{R}^{6}B_{g}^{3} \overline{Ric}  -68864 \overline{Ric}^{2}{\pi }^{2}{R}^{4}B_{g}^{2}\nonumber\\-480  \overline{Ric}  ^{3}\pi {R}^{2}B_{g}+225 \overline{Ric} ^{4} \Bigg\}^{1/2}\Bigg].
	     \end{align}  
     }

\section{Finsler metric in Randers Space}
 There are two types of space noted in Finslerian metric, (i) Riemannian space-time and (ii) Randers space-time. Randers space-time can be written in the form 
\begin{equation}
 F(x,y) \equiv \alpha(x,y) +\beta(x,y),
\end{equation}
where $\alpha$ and $\beta$ are the Riemannina metric and 1-form respectively.
 
  \[
  \alpha(x,y) =\sqrt{a_{\mu \nu}(x) y^\mu y^\nu}, ~~\beta(x,y)= b_\mu(x) y^\mu. 
  \]

For constant $\overline{Ric}$ ($=\chi$), the Finsler structure $\overline{F}^2$ will be
\begin{align}
	\overline{F}^2 &=y^\theta y^\theta + X \sin^2(\sqrt{\chi} \theta) y^\phi y^\phi,~~~\chi>0, \label{c1}\\
	&=y^\theta y^\theta + X  \theta^2 y^\phi y^\phi,~~~\chi=0, \label{c2}\\
	&=y^\theta y^\theta + X \sinh^2(\sqrt{-\chi} \theta) y^\phi y^\phi,~~~\chi<0. \label{c3}
\end{align}	
where Eq. (\ref{c1}) refers to the space that are both forward and backward geodesically comeplete must be Riemannian, the space of Eq. (\ref{c2}) is locally Minkowskian and last one Eq. (\ref{c3}) is least understood~\cite{Bao2000}. 

Now, the Finsler structure F$^2$ takes the form with Eq. (\ref{c1}), is as follows
\begin{align}
F^2 &= e^{\lambda(r)} y^t y^t - e^{\nu(r)} y^r y^r -r^2 (y^\theta y^\theta +  \sin^2(\sqrt{\chi} \theta) y^\phi y^\phi ) \nonumber\\
&= \alpha^2+r^2\psi(\theta) y^\phi y^\phi,
\end{align}
where $\psi(\theta)= \sin^2 \theta - \sin^2(\sqrt{\chi}\theta)$ with $ X $=1.

Hence
\begin{equation}
F= \alpha\sqrt{1+\frac{(b_\phi y^\phi)^2}{\alpha^2}}, \label{c5}
\end{equation}
where \[b_\phi =r \sqrt{\psi(\theta)}\].

Eq. (\ref{c5}) can be rewriten as
\begin{equation}
F=\alpha \sqrt{1+s^2}.
\end{equation}

This is the Finsler metric in Randers space with the choice $s = \frac{b_\phi y^\phi}{\alpha} = \frac{\beta}{\alpha}$ where
\[
b_\mu=(0,0,0,b_\phi).
\]

By using Eq. (\ref{c1}), the volume of closed Finsler-Randers suface can be written as 
\begin{equation}
Vol_F= \int \left(1-(a^{ij}b_ib_j)^{3/2}\right) \sqrt{det(a_{ij})}dx^1 \wedge  dx^2.
\end{equation}

Li and Chang~\cite{Li2014} found the volume as 4$\pi$, same as in Riemannian sphere.\\

\section*{ACKNOWLEDGMENTS}
SR and FR are thankful to the Inter-University Centre for Astronomy and Astrophysics (IUCAA), Pune, India for providing Visiting Associateship under which a part of this work was carried out. SR is also thankful to the authority of The Institute of Mathematical Sciences, Chennai, India and the Centre for Theoretical Studies, IIT Kharagpur, India for providing short term visits under which a part of this work was carried out. FR is also grateful to DST-SERB (EMR/2016/000193), Government of India for providing financial support. A part of this work was completed while SRC and DD were visiting IUCAA and the authors gratefully acknowledge the warm hospitality and facilities there.

\end{document}